\documentclass[]{elsarticle}

\usepackage{lineno,hyperref}
\usepackage{epsfig}
\usepackage{graphicx}
\usepackage{dcolumn}
\usepackage{amsmath}
\usepackage{bm}
\modulolinenumbers[5]

\journal{}

\begin{document}

\begin{frontmatter}

\title{From minijet saturation to global observables in $A$ + $A$ collisions at the LHC and RHIC}

\author[JKL,HIP]{R. Paatelainen}
\author[JKL,HIP]{K.~J. Eskola}
\author[JKL,HIP]{H. Niemi}
\author[HKI,HIP]{K.~ Tuominen}
\address[JKL]{Department of Physics,  University of Jyv\"{a}skyl\"{a},\\ P.O.Box 35, FI-40014 University of Jyv\"{a}skyl\"{a}, Finland}
\address[HIP]{Helsinki Institute of Physics,\\ P.O.Box 64, FI-00014 University of Helsinki, Finland}
\address[HKI]{Department of Physics, University of Helsinki,\\ P.O.Box 64, FI-00014 University of Helsinki, Finland}

\begin{abstract}

We review the recent results from the computation of saturated next-to-leading order perturbative QCD minijet intial conditions combined with viscous hydrodynamical evolution of ultrarelativistic heavy-ion collisions at the LHC and RHIC. Comparison with experimental data is shown.
\end{abstract}


\end{frontmatter}

\section{Introduction}

In this proceedings we report the results from our recent study \cite{Paatelainen:2013at}, where we have computed the local initial energy densities and formation times of the produced quark-gluon plasma (QGP) for the fluid-dynamical evolution in ultrarelativistic heavy-ion collisions at the LHC and RHIC. The basis of the framework is a rigorous next-to-leading order (NLO) perturbative QCD (pQCD) minijet transverse energy $(E_T)$ calculation using the latest nuclear parton distribution functions \cite{EPS09,EPS09s}. The production of the initial energy is then assumed to be moderated by gluon saturation. 

By using viscous hydrodynamics, we show that we can obtain a good simultaneous description of the centrality dependence of multiplicity, transverse momentum $(p_T)$ spectra and elliptic flow $(v_2)$ measured 
in Au+Au collisions at RHIC and Pb+Pb collisions at the LHC. In particular, the shear viscosity in the different phases of QCD matter is constrained in this framework simultaneously by all these data.

\section{Model setup and results}

The saturation criterion for the minijet $E_T$ production in $A$+$A$ collision at non-zero impact parameters, is formulated as \cite{Paatelainen:2013at} 
\begin{equation}
\label{eq: saturation}
\frac{{\rm d}E_T}{{\rm d}^2\mathbf{s}}(p_0,\sqrt{s_{{\rm NN}}},\Delta y,\mathbf{s},\mathbf{b},\beta) = \frac{K_{\rm sat}}{\pi}p_0^3\Delta y,
\end{equation}
where ${\rm d}E_T/{\rm d}^2\mathbf{s}$ is the computed minijet $E_T$ above a $p_T$ scale $p_0$, produced into a rapidity region $\Delta y=1$ at an impact parameter $\mathbf{b}$. Here $K_{\rm sat}$ is an unknown proportionality constant of the order of $\mathcal{O}(1)$, $\mathbf{s}$ is the transverse position and $\sqrt{s_{NN}}$ is the cms-energy in the collision. As discussed in \cite{Paatelainen:2013at, Paatelainen:2012at}, the parameter $\beta \in [0,1]$ controls the soft pQCD kinematics inside the acceptance window $\Delta y$. Once the solution $p_0=p_{\rm sat}$ of the transversally local saturation criterion,  Eq.\ \eqref{eq: saturation}, is known for given $K_{\rm sat}$ and $\beta$, the local energy density is obtained as 
\begin{equation}
\epsilon(\mathbf{s},\tau_s) = \frac{{\rm d}E_T}{{\rm d}^2\mathbf{s}\tau_s\Delta y} = \frac{K_{\rm sat}}{\pi}p_{\rm sat}^4,
\end{equation}
where the local formation time is $\tau_s = 1/p_{\rm sat}$. Note that the formation time $\tau_s$ is different at different points in the transverse plane. However, for the hydrodynamical evolution, we need the initial state at a fixed $\tau_0$. The prethermal evolution from $\tau_s$ to $\tau_0 = \frac{1}{1~{\rm GeV}}\simeq 0.2 ~{\rm fm}$ is obtaind using either Bjorken free streaming (FS) or the Bjorken hydrodynamic scaling solution (BJ). For more details see Ref.\ \cite{Paatelainen:2013at}.

\begin{figure}[!]
\centering
\epsfxsize 5.9cm \epsfbox{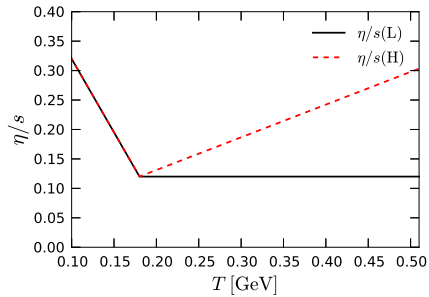}  
\caption{\protect (Color online)
The parametrizations "L" and "H" of the shear viscosity to entropy density ratio. From \cite{Paatelainen:2013at}.
}
\label{fig:results3}
\end{figure}

For the  hydrodynamical evolution, we take the 2+1 D setup introduced in \cite{Niemi:2011ix}. We use the lattice QCD and hadron resonance gas based equation of state s95p-PCE-v1 \cite{Huovinen:2009yb} with a chemical freeze-out temperature $T_{\rm chem} = 175$ MeV. The kinetic freeze-out temperature is here always $T_{\rm dec} =100$ MeV. The parametrizations of the temperature-dependent shear viscosity to entropy ratio $\eta/s(T)$, for which we show the following results, are shown in Fig.\ \ref{fig:results3}. The shear-stress and transverse flow are initially set to zero.

\begin{figure*}[!]
\epsfxsize 5.9cm \epsfbox{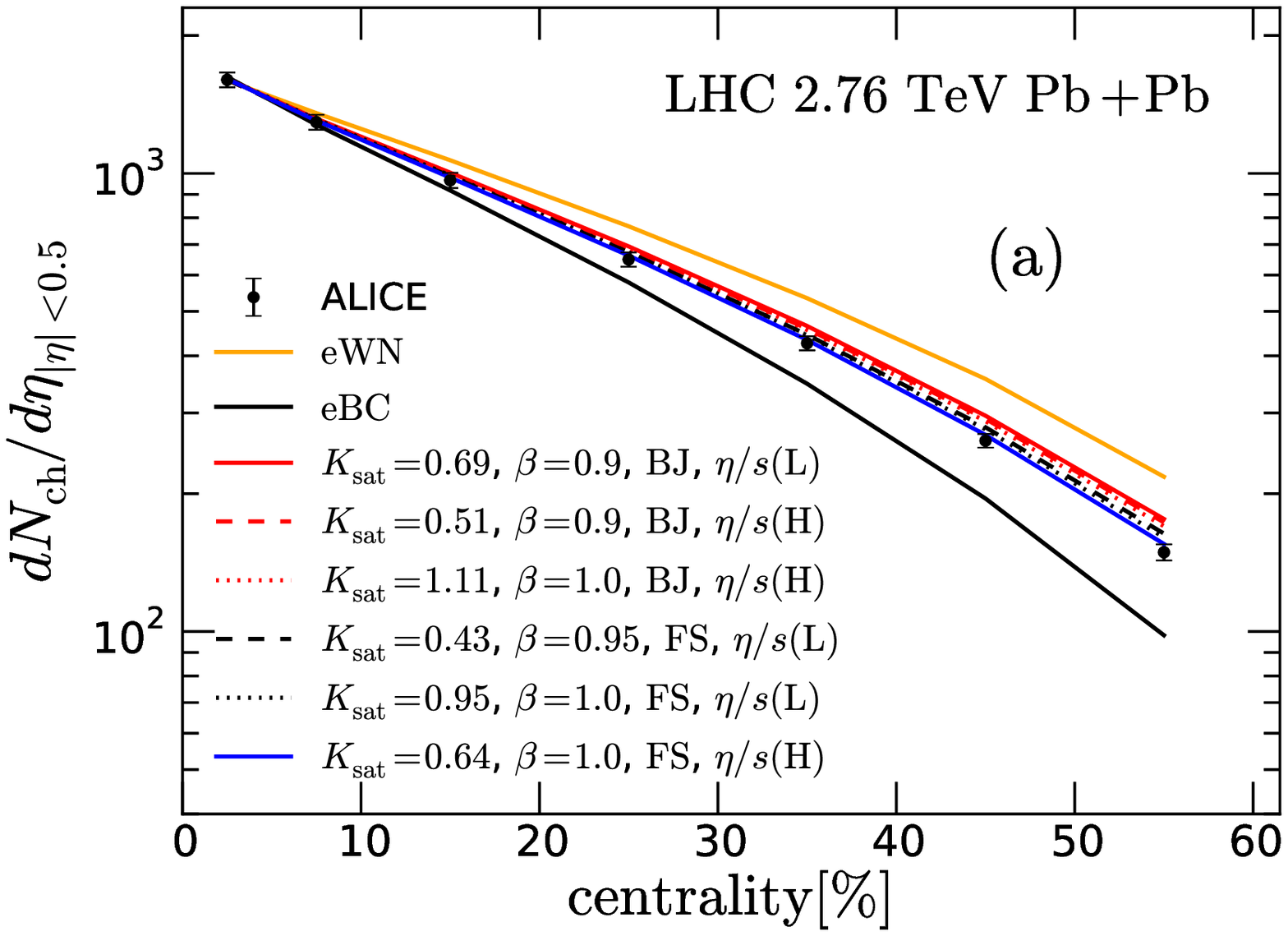} 
\epsfxsize 5.9cm \epsfbox{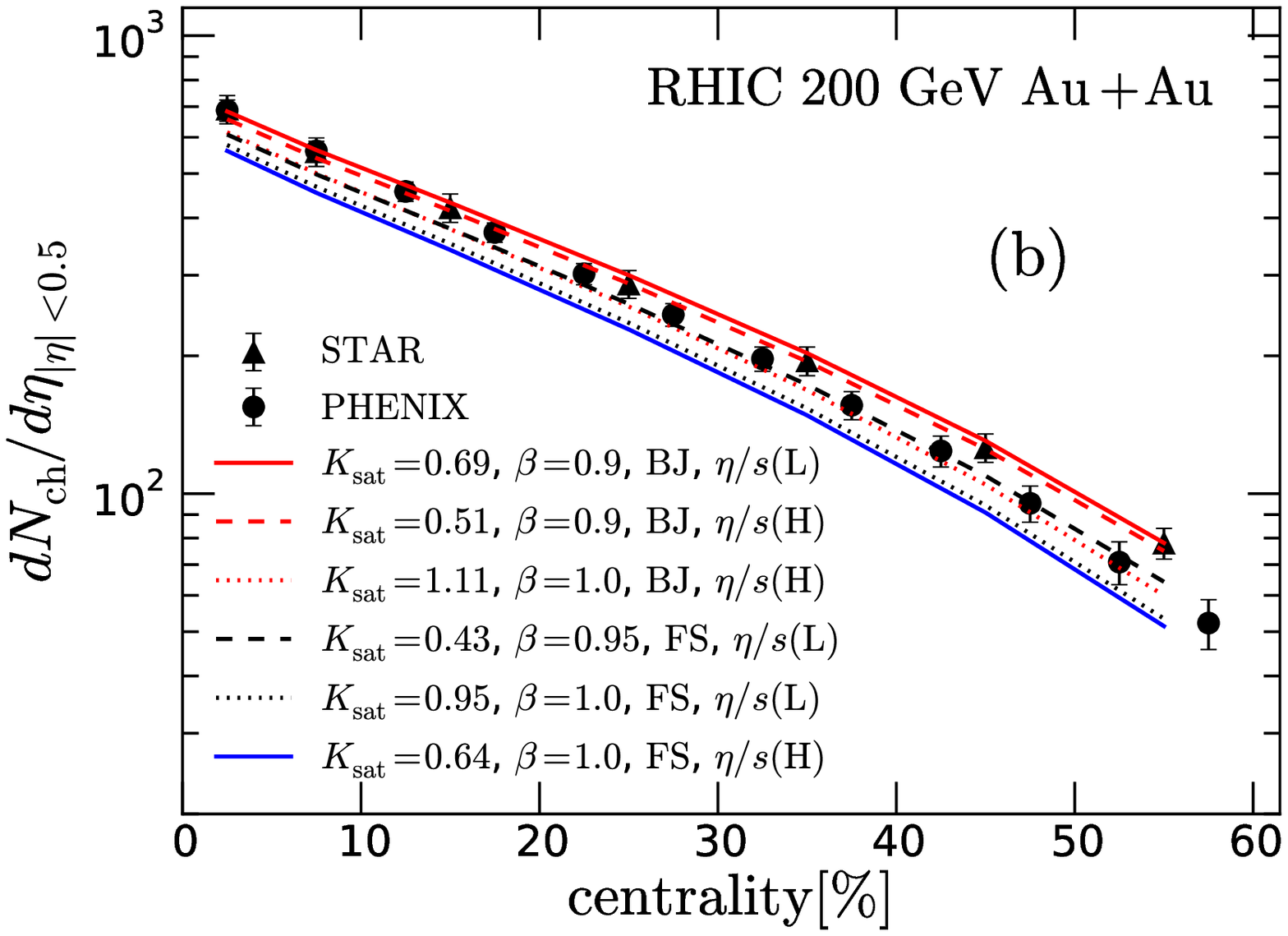} 
\epsfxsize 5.9cm \epsfbox{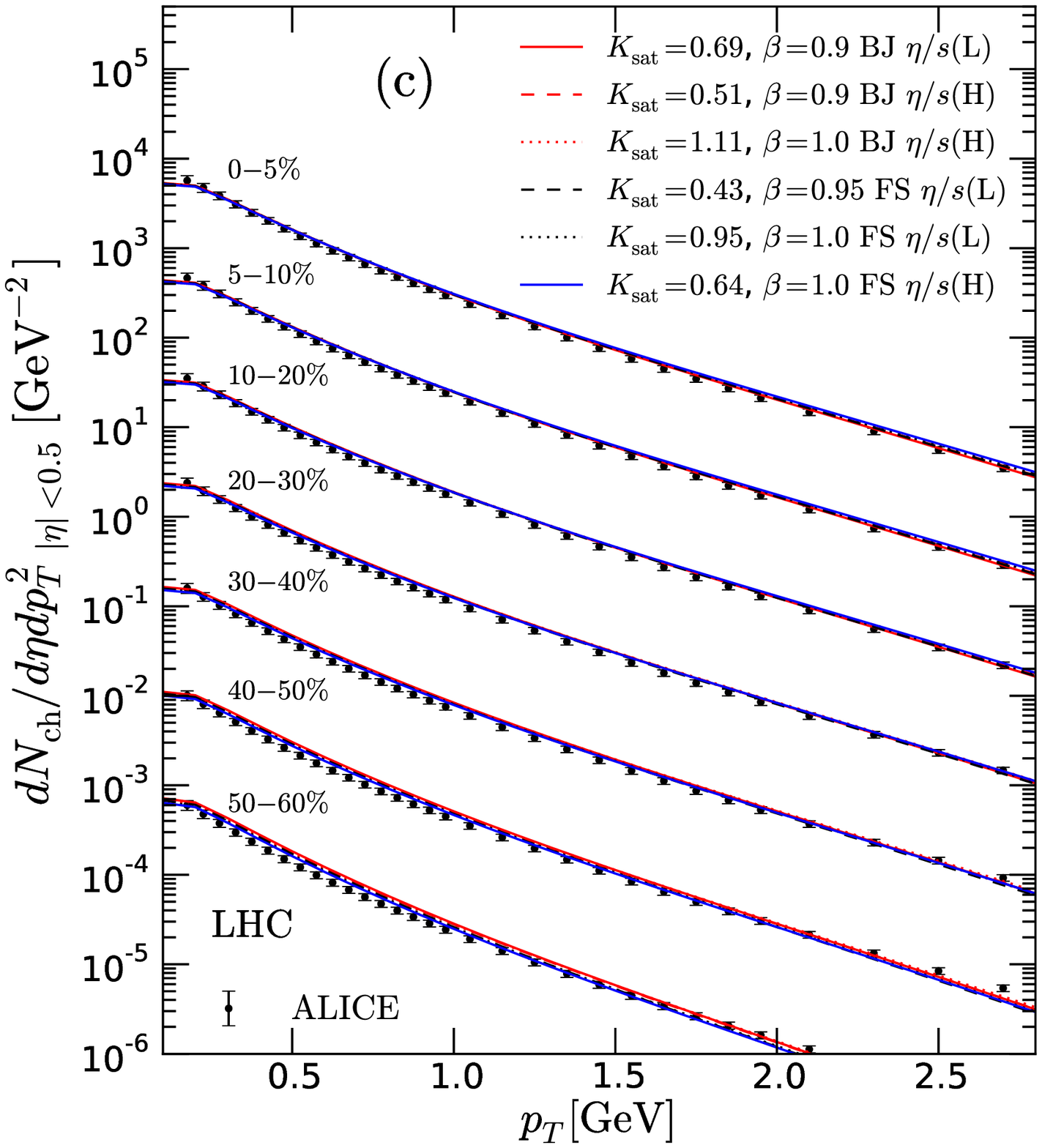} 
\epsfxsize 5.9cm \epsfbox{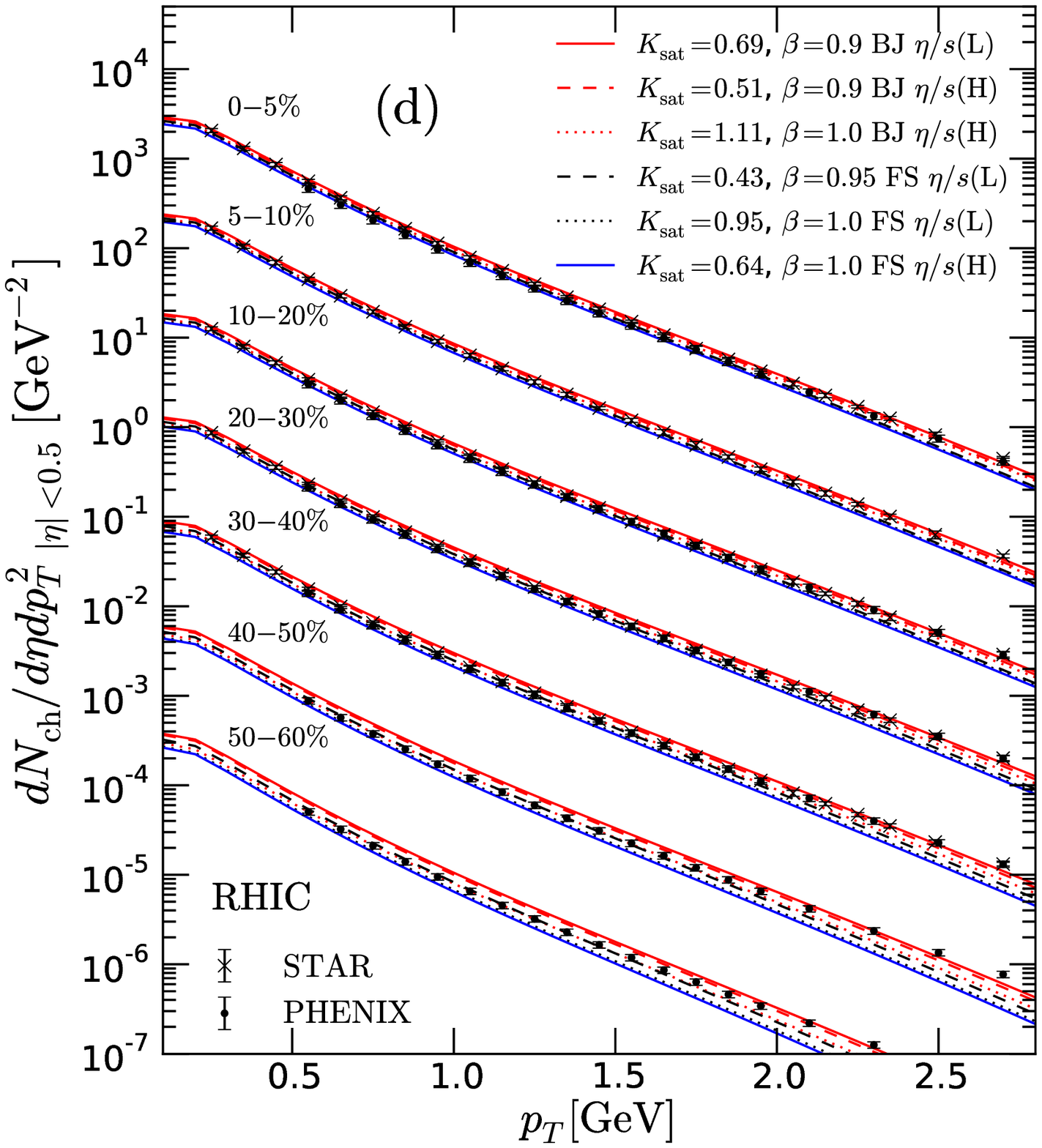} 
\epsfxsize 5.9cm \epsfbox{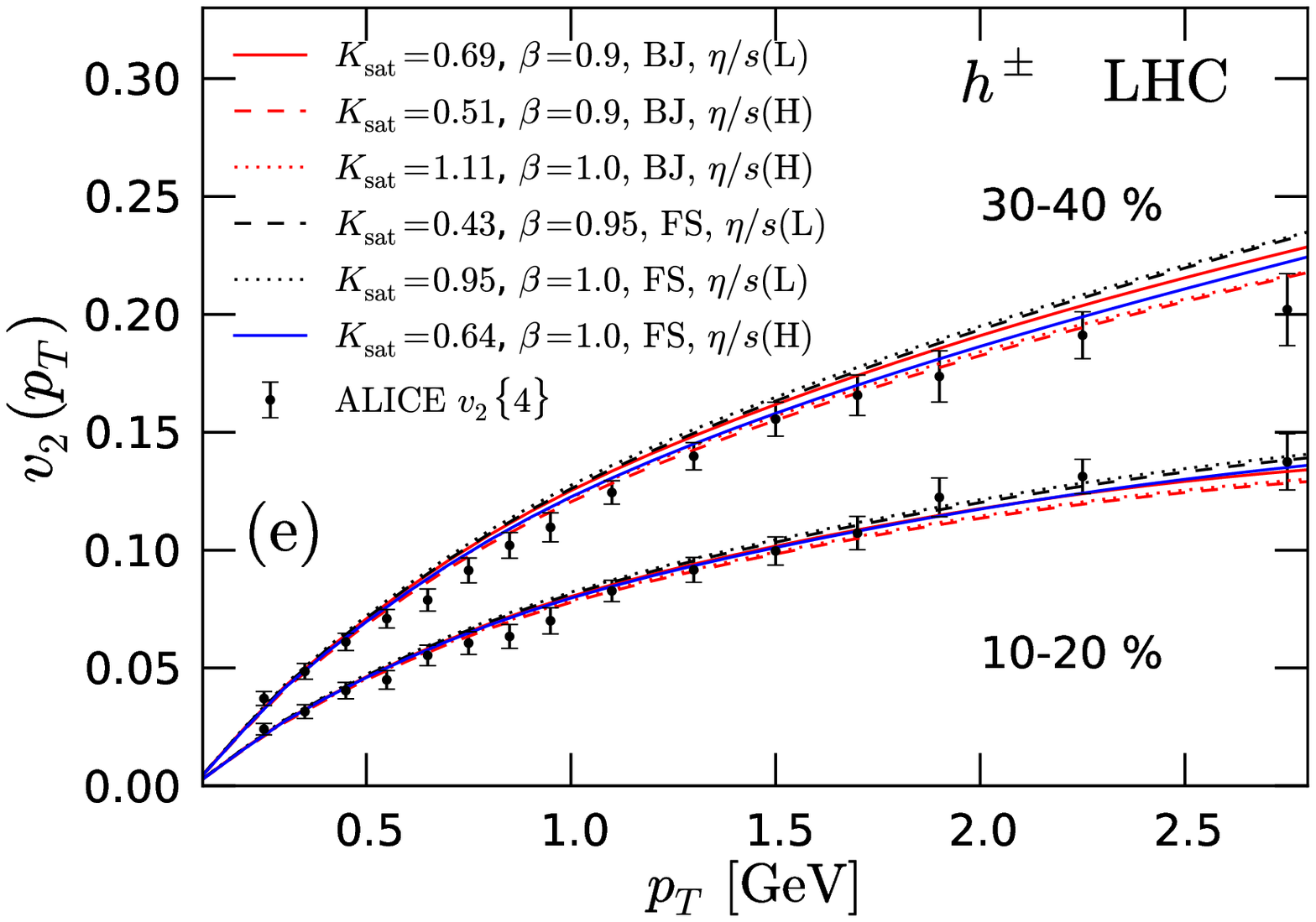} 
\epsfxsize 5.9cm \epsfbox{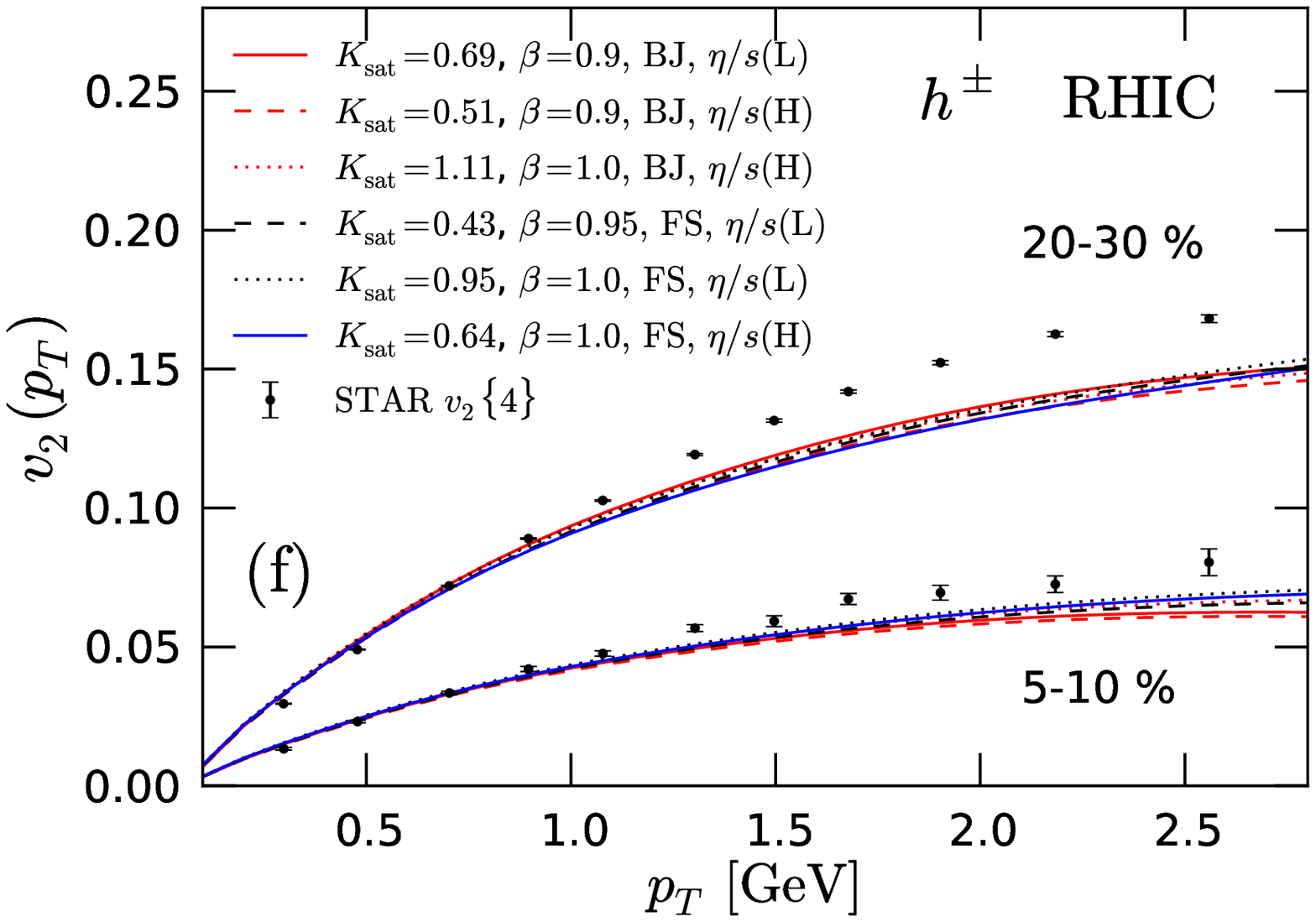} 
\caption{\protect (Color online)
Centrality dependence of the charged hadron multiplicity at the LHC (a) and RHIC (b). 
$p_T$ spectra of charged hadrons at the LHC (c) and RHIC (d), in the same centrality classes as the ALICE data in panel (a), and scaled down by increasing powers of 10.
Elliptic flow coefficients $v_2(p_T)$ at the LHC (e) and RHIC (f), compared with the measured 4-particle cumulant $v_2\{4\}(p_T)$. Labeling of the theory curves in each panel is identical, and the parameter sets $\{K_{\rm sat}, \beta, {\rm BJ/FS}, \eta/s(T) \}$ are indicated. The labels H and L refer to Fig.~\ref{fig:results3}. From \cite{Paatelainen:2013at}.
}
\label{fig: results4}
\end{figure*}

In Fig.\ \ref{fig: results4}a and \ref{fig: results4}b we show the computed centrality dependence of the charged hadron multiplicity in Pb+Pb collisions at $\sqrt{s_{NN}}=2.76$ TeV and in Au+Au collisions at $\sqrt{s_{NN}}=200$ GeV compared with the ALICE \cite{Aamodt:2010pb}, PHENIX \cite{PHENIXMUL:2005} and STAR \cite {STARSPECT:2009} data. For a given set of parameters $\{\beta, {\rm BJ}/{\rm FS}, \eta/s(T)\}$ the remaining parameter $K_{\rm sat}$ is tuned such that the multiplicity in the 0-5\%  most central collisions at the LHC is reproduced. Next, the obtained centrality dependence of the computed $p_T$-spectra of charged hadrons are shown in Fig. \ref{fig: results4}c for the LHC and in Fig.\ \ref{fig: results4}d for RHIC. The data are from \cite{Abelev:2012hxa} and \cite{Adams:2003kv,Adler:2003au}, correspondingly. Finally, in Figs.\ \ref{fig: results4}e and \ref{fig: results4}f we show the elliptic flow coefficients $v_2(p_T)$ at the LHC and RHIC, respectively. The data are from \cite{Aamodt:2010pa} (ALICE) and \cite{Bai} (STAR).

To conclude, we note the following: the $p_T$ spectra are not very sensitive to the parameters $\{\beta, {\rm BJ}/{\rm FS}, \eta/s(T)\}$ once the centrality dependence of the multiplicities is under control. The $v_n$ coefficients, however, depend strongly on the $\eta/s(T)$ parametrization: an ideal fluid description would fail to reproduce the measured $v_2(p_T)$, while with both the "L" and "H" parametrizations we get a good agreement with the data. Before entering a more complete global analysis for $\eta/s(T)$, the initial event-by-event density fluctuations need to be considered in this framework. This is a work in progress.

\vspace{0.1cm}

\textbf{Acknowledgments}: This work was financially supported by the Jenny and Antti Wihuri Foundation (RP) and 
the Academy projects of Finland 133005 (KJE) and 267842 (KT). We thank CSC-IT Center Science for supercomputing time.


\end{document}